\documentclass{PoS}

\title{\vspace*{-2mm} {\tiny DESY 10--122 \hfill SFB/CPP-10-74 \hfill TTK-10-40 \hfill IFIC-10-23}
\\
\boldmath Logarithmic $O(\alpha_s^3)$ contributions to the DIS Heavy Flavor Wilson 
Coefficients 
at $Q^2 \gg m^2$}

\ShortTitle{Logarithmic $O(\alpha_s^3)$ contributions}

\author{Isabella Bierenbaum\thanks{Supported in part by Generalitat Valenciana  Grant No. 
PROMETEO/2008/69.}\\
        Instituto de F\'{i}sica Corpuscular, CSIC-Universitat de Val\`{e}ncia, Apartado de Correos 
22085, 
E--46071 Valencia, Spain\\
        E-mail: \email{Isabella.Bierenbaum@ific.uv.es}}

\author{\speaker{Johannes Bl\"umlein}\thanks{Supported in part by SFB-TR/9 and EU TMR network 
HEPTOOLS.}\\
        Deutsches Elektronen--Synchrotron, DESY, Platanenallee 6, D--15638 Zeuthen, Germany\\
        E-mail: \email{Johannes.Bluemlein@desy.de}}

\author{Sebastian Klein\thanks{Supported in part by SFB-TR/9.}\\
        Institut f\"ur theoretische Teilchenphysik und Kosmologie, RWTH Aachen University, D--52056 
Aachen, Germany\\
        E-mail: \email{sklein@physik.rwth-aachen.de}}

\abstract{The logarithmic contributions to the massive twist-2 operator matrix elements for 
deep-inelastic scattering are calculated to $O(\alpha_s^3)$for general values of the Mellin 
variable $N$.}

\FullConference{XVIII International Workshop on Deep-Inelastic Scattering and Related Subjects, DIS 2010\\
		April 19-23, 2010\\
		Firenze, Italy}

\newcommand{\N}{\nonumber}

\newcommand{\Ctil}{\tilde{C}}

\begin{document}
\section{Introduction}

\noindent
The heavy flavor contributions to deep-inelastic scattering (DIS) are rather large in the small $x$ 
region.
With the current DIS data a precision of $\sim 1\%$ is reached~\cite{H1Z:2009wt} for $F_2(x,Q^2)$. This 
requires to describe the heavy flavor corrections to 3-loop order, to perform a consistent next-to-next-to 
leading order (NNLO) analysis, to measure the strong coupling constant $\alpha_s(M_Z^2)$ and the 
unpolarized twist-2 parton distribution functions at highest precision possible,~cf.~\cite{PDF}.
In the region $Q^2/m^2 \geq 10$ one may compute all contributions except 
the power suppressed terms, $\propto (m^2/Q^2)^k, k\geq 1$ using the factorization 
theorem given in Ref.~\cite{Buza:1995ie}. Here, the massive Wilson coefficients factorize into massive
operator matrix elements (OMEs), $A_{ij}$, and the massless Wilson coefficients, which are known to 
$O(\alpha_s^3)$~\cite{Vermaseren:2005qc}. The $O(\alpha_s^3)$ Mellin moments of the massive operator matrix 
elements up to $N = 10 ... 14$, depending on the process, were computed in \cite{Bierenbaum:2009mv}.
The calculation was performed relating the moments of the massive operator matrix elements    
to massive tadpoles and using {\tt MATAD}~\cite{Steinhauser:2000ry}.
In Ref.~\cite{Bierenbaum:2009mv} also the complete renormalization for a single massive quark has been 
derived. Different 
other contributions, needed in the renormalization process, were computed at general values of $N$ in 
Refs.~\cite{TOL}. For the structure 
function $F_L(x,Q^2)$ the asymptotic corrections to $O(\alpha_s^3)$ are known for general 
values of $N$~\cite{Blumlein:2006mh}. They are, however, only applicable at scales $Q^2/m^2 \geq 800$.
For transversity the matrix elements were computed for general $N$ at 2-loop order and a series of
moments at 3-loop order in \cite{Blumlein:2009rg}. 
Very recently, the general $N$ results at 
$O(\alpha_s^3)$ for $F_2(x,Q^2)$ for the color coefficients $\propto n_f$ have been 
completed~\cite{Ablinger1,Ablinger:2010ha}. These computations use modern summation technologies encoded
in the package {\tt Sigma}~\cite{SIGMA} and the results can be expressed in terms of nested harmonic sums
\cite{HSUM}. 
From the single pole terms in the massive computations of
Refs.~\cite{Bierenbaum:2009mv,Blumlein:2009rg,Ablinger1,Ablinger:2010ha}
the corresponding contributions to the 3--loop anomalous dimensions were derived, either for the respective 
moments \cite{ANDIM1}, or at general values of $N$~\cite{ANDIM2}.

In this note we report on the logarithmic $O(\alpha_s^3)$ contributions to the massive operator matrix 
elements,~cf. also \cite{BBK1}. They are known for general values of $N$ and depend on the 3-loop anomalous 
dimensions and massive OMEs up to $O(\alpha_s^2)$. 
\section{The heavy flavor Wilson coefficients in the asymtotic region}

\noindent
The heavy flavor correction to the structure function $F_2(x,Q^2)$ with $n_f$ massless and one
heavy flavor reads,~\cite{Bierenbaum:2009mv}~:
\begin{eqnarray}
      \label{eqF2}
       F_{(2,L)}^{Q\overline{Q}}(x,n_f,Q^2,m^2) &=&
       \sum_{k=1}^{n_f}e_k^2\Biggl\{
                   L_{q,(2,L)}^{\sf NS}\left(x,n_f+1,\frac{Q^2}{m^2}
                                                ,\frac{m^2}{\mu^2}\right)
                \otimes
                   \Bigl[f_k(x,\mu^2,n_f)+f_{\overline{k}}(x,\mu^2,n_f)\Bigr]
\N\\ &&\hspace{-35mm}
               +\frac{1}{n_f}\Bigl[L_{q,(2,L)}^{\sf PS}\left(x,n_f+1,\frac{Q^2}{m^2}
                                                ,\frac{m^2}{\mu^2}\right)
                \otimes
                   \Sigma(x,\mu^2,n_f)
               + L_{g,(2,L)}^{\sf S}\left(x,n_f+1,\frac{Q^2}{m^2}
                                                 ,\frac{m^2}{\mu^2}\right)
                \otimes
                   G(x,\mu^2,n_f)
                             \Bigr] \Biggr\}
\N\\ &&\hspace{-35mm}
+ e_Q^2\Biggl[
                   H_{q,(2,L)}^{\sf PS}\left(x,n_f+1,\frac{Q^2}{m^2}
                                        ,\frac{m^2}{\mu^2}\right)
                \otimes
                   \Sigma(x,\mu^2,n_f)
                  +H_{g,(2,L)}^{\sf S}\left(x,n_f+1,\frac{Q^2}{m^2}
                                           ,\frac{m^2}{\mu^2}\right)
                \otimes
                   G(x,\mu^2,n_f)
                                  \Biggr]~,\N\\
\end{eqnarray}
with boundaries for the Mellin integral $[x(1+4m^2/Q^2), 1]$, and $e_i$ the quark charges.
Here the different Wilson coefficients are denoted by $L_i, H_i$ in case the photon couples to
a light (L) or the heavy (H) quark. For $Q^2 \gg m^2$ they can be expressed in terms of the
massive OMEs $A_{ij}$ and the massless Wilson coefficients $C_j$. To $O(a_s^3)$ they read ($a_s = 
\alpha_s/(4\pi))$
\begin{eqnarray}
\label{eqWIL1}
     L_{q,(2,L)}^{\sf NS}(n_f+1) &=&
     a_s^2 \Bigl[A_{qq,Q}^{(2), {\sf NS}}(n_f+1) \delta_2 +
     \hat{C}^{(2), {\sf NS}}_{q,(2,L)}(n_f)\Bigr]
     \N\\
     &+&
     a_s^3 \Bigl[A_{qq,Q}^{(3), {\sf NS}}(n_f+1) \delta_2
     +  A_{qq,Q}^{(2), {\sf NS}}(n_f+1) C_{q,(2,L)}^{(1), {\sf NS}}(n_f+1)
     + \hat{C}^{(3), {\sf NS}}_{q,(2,L)}(n_f)\Bigr]  
      \label{eqWIL2}
\nonumber\\
      L_{q,(2,L)}^{\sf PS}(n_f+1) &=&
     a_s^3 \Bigl[~A_{qq,Q}^{(3), {\sf PS}}(n_f+1)~\delta_2
     +  A_{gq,Q}^{(2)}(n_f)~~n_f\Ctil_{g,(2,L)}^{(1)}(n_f+1) 
     + n_f \hat{\Ctil}^{(3), {\sf PS}}_{q,(2,L)}(n_f)\Bigr]
         \label{eqWIL3}
\nonumber\\
      L_{g,(2,L)}^{\sf S}(n_f+1) &=&
     a_s^2 A_{gg,Q}^{(1)}(n_f+1)n_f \Ctil_{g,(2,L)}^{(1)}(n_f+1)
+
      a_s^3 \Bigl[~A_{qg,Q}^{(3)}(n_f+1)~\delta_2 \N\\ &&
     +  A_{gg,Q}^{(1)}(n_f+1)~~n_f\Ctil_{g,(2,L)}^{(2)}(n_f+1)
     +  A_{gg,Q}^{(2)}(n_f+1)~~n_f\Ctil_{g,(2,L)}^{(1)}(n_f+1)
     \N\\ && 
     +  ~A^{(1)}_{Qg}(n_f+1)~~n_f\Ctil_{q,(2,L)}^{(2), {\sf PS}}(n_f+1)
     + n_f \hat{\Ctil}^{(3)}_{g,(2,L)}(n_f)\Bigr]~,
\nonumber\\
     H_{q,(2,L)}^{\sf PS}(n_f+1)
     &=& a_s^2 \Bigl[~A_{Qq}^{(2), {\sf PS}}(n_f+1)~\delta_2
     +~\Ctil_{q,(2,L)}^{(2), {\sf PS}}(n_f+1)\Bigr]
+ a_s^3 \Bigl[~A_{Qq}^{(3), {\sf PS}}(n_f+1)~\delta_2 
\N\\ &&
     +~\Ctil_{q,(2,L)}^{(3), {\sf PS}}(n_f+1) 
     + A_{gq,Q}^{(2)}(n_f+1)~\Ctil_{g,(2,L)}^{(1)}(n_f+1)
     \N\\ &&
+ A_{Qq}^{(2), {\sf PS}}(n_f+1)~C_{q,(2,L)}^{(1), {\sf NS}}(n_f+1)
        \Bigr]~,       \label{eqWIL4}
\nonumber\\
     H_{g,(2,L)}^{\sf S}(n_f+1) &=& a_s \Bigl[~A_{Qg}^{(1)}(n_f+1)~\delta_2
     +~\Ctil^{(1)}_{g,(2,L)}(n_f+1) \Bigr]   + a_s^2 \Bigl[~A_{Qg}^{(2)}(n_f+1)~\delta_2
\N\\ &&
     +~A_{Qg}^{(1)}(n_f+1)~C^{(1), {\sf NS}}_{q,(2,L)}(n_f+1)
     +~A_{gg,Q}^{(1)}(n_f+1)~\Ctil^{(1)}_{g,(2,L)}(n_f+1)
     \N\\ &&
     +~\Ctil^{(2)}_{g,(2,L)}(n_f+1) \Bigr]
     +~a_s^3 \Bigl[~A_{Qg}^{(3)}(n_f+1)~\delta_2
     +~A_{Qg}^{(2)}(n_f+1)~C^{(1), {\sf NS}}_{q,(2,L)}(n_f+1)
     \N\\ &&
     +~A_{gg,Q}^{(2)}(n_f+1)~\Ctil^{(1)}_{g,(2,L)}(n_f+1)
     +~A_{Qg}^{(1)}(n_f+1)\Bigl\{
     C^{(2), {\sf NS}}_{q,(2,L)}(n_f+1)
     \N\\ && 
     +~\Ctil^{(2), {\sf PS}}_{q,(2,L)}(n_f+1)\Bigr\}
     +~A_{gg,Q}^{(1)}(n_f+1)~\Ctil^{(2)}_{g,(2,L)}(n_f+1)
     +~\Ctil^{(3)}_{g,(2,L)}(n_f+1) \Bigr]~, 
\N\\
\label{eqWIL5}
\end{eqnarray}
with $\delta_2 = 0 (1)$ for $F_L (F_2)$ and $\hat{f}(n_f) = f(n_f+1) - f(n_f), \tilde{f}(n_f) = 
f(n_f)/n_f$.
The massive OMEs depend on the ratio $m^2/\mu^2$, while the scale ratio of the massless 
Wilson coefficients is $\mu^2/Q^2$. The latter are pure functions of the momentum fraction $z$, or the
Mellin variable $N$, if one sets $\mu^2 = Q^2$. The massive OMEs obey then the general structure
\begin{eqnarray}
\label{eq:log}
A^{(3)}_{ij}\left(\frac{m^2}{Q^2}\right)  = 
  a^{(3),3}_{ij} \ln^3\left(\frac{m^2}{Q^2}\right)
+ a^{(3),2}_{ij} \ln^2\left(\frac{m^2}{Q^2}\right)
+ a^{(3),1}_{ij} \ln\left(\frac{m^2}{Q^2}\right)
+ a^{(3),0}_{ij}~.
\end{eqnarray}
\section{The matrix element $A_{Qg}^{(3)}(N)$}

\noindent
In the following we present, as an example, the logarithmic expansion coefficients of Eq.~(\ref{eq:log}), 
$a^{(3),k}, k \geq 1$, for the massive OME $A_{Qg}^{(3)}(N)$ in the $\overline{\rm MS}$--scheme. They are 
given by~:
%
%
{\small
 \begin{eqnarray}
   a_{Qg}^{(3),3}&=&\frac{8(N^2+N+2)T_F}{9N(N+1)(N+2)}\Biggl[
     T_Fn_f\Biggl(
          C_F\Biggl(
                \frac{P_1}{(N-1)N^2(N+1)^2(N+2)}
               -4S_{1}
             \Biggr)
 \N\\ && \hspace{-10mm}
         +C_A\Biggl(
                4S_{1}
               -\frac{8(N^2+N+1)}{(N-1)N(N+1)(N+2)}
             \Biggr)
           \Biggr)
    -8T_F^2
    +C_A^2 \Biggl(
             -\frac{(11N^4+22N^3-59N^2-70N-48)S_{1}}{(N-1)N(N+1)(N+2)}
 \N\\ && \hspace{-10mm}
             -12S_{1}^2
             +\frac{2(N^2+N+1)(11N^4+22N^3-35N^2-46N-24)}{(N-1)^2N^2(N+1)^2(N+2)^2}
           \Biggr)
    +C_AT_F\Biggl(
             -\frac{56(N^2+N+1)}{(N-1)N(N+1)(N+2)}
 \N\\ && \hspace{-10mm}
             +28S_{1}
           \Biggr)
    +C_F^2\Biggl(
             -3\frac{(3N^2+3N+2)^2}{4N^2(N+1)^2}
             +\frac{6S_{1}(3N^2+3N+2)}{N(N+1)}
             -12S_{1}^2
          \Biggr)
    +C_FT_F\Biggl(
             -16S_{1}
 \N\\ && \hspace{-10mm}
             +\frac{2P_2}{(N-1)N^2(N+1)^2(N+2)}
           \Biggr)
    +C_AC_F\Biggl(
              24S_{1}^2
             -\frac{(N^2+N+6)(7N^2+7N+4)S_{1}}{(N-1)N(N+1)(N+2)}
 \N\\ && \hspace{-10mm}
             -\frac{(3N^2+3N+2)(11N^4+22N^3-59N^2-70N-48)}{4(N-1)N^2(N+1)^2(N+2)}
          \Biggr)
                                                    \Biggr]~.\label{AQg33}
 \end{eqnarray}
 }
%
%
{\small
 \begin{eqnarray}
  a_{Qg}^{(3),2}&=&4T_F^2n_f\Biggl(
    C_F\Biggl(
         -\frac{4(N^2+N+2)}{3N(N+1)(N+2)}\Bigl(S_{2}+S_{1}^2\Bigr)
         +\frac{8(5N^3+8N^2+19N+6)S_{1}}{9N^2(N+1)(N+2)}
 \N\\ && \hspace{-10mm}
         -\frac{P_3}{9(N-1)N^4(N+1)^4(N+2)^3}
       \Biggr)
   +C_A\Biggl(
         -\frac{8(5N^4+20N^3+47N^2+58N+20)S_{1}}{9N(N+1)^2(N+2)^2}
 \N\\ && \hspace{-10mm}
         +\frac{4(N^2+N+2)}{3N(N+1)(N+2)}\Bigl(2S_{-2}+S_{2}+S_{1}^2\Bigr)
         -\frac{2P_4}{9(N-1)N^2(N+1)^3(N+2)^3}
       \Biggr)
                         \Biggr)
  +2C_A^2T_F\Biggl(
 \N\\ && \hspace{-10mm}
               \frac{8(N^2+N+2)}{N(N+1)(N+2)}
                \Bigl(2S_{-2,1}-S_{1}^3-3S_{2}S_{1}-S_{-3}-4S_{-2}S_{1}-S_{3}\Bigr)
              -\frac{2P_5S_{1}^2}{3(N-1)N^2(N+1)^2(N+2)^2}
 \N\\ && \hspace{-10mm}
              +\frac{4P_6S_{1}}{9(N-1)^2N^3(N+1)^3(N+2)^3}
              -\frac{4(N^2+N+2)(11N^4+22N^3-59N^2-70N-48)S_{-2}}{3(N-1)N^2(N+1)^2(N+2)^2}
 \N\\ && \hspace{-10mm}
              +\frac{8P_7}{9(N-1)^2N^4(N+1)^4(N+2)^4}
              -\frac{2(N^2+N+2)(11N^4+22N^3-83N^2-94N-72)S_{2}}{3(N-1)N^2(N+1)^2(N+2)^2}
            \Biggr)
 \N\\ && \hspace{-10mm}
  +4C_AT_F^2\Biggl(
               \frac{4(N^2+N+2)}{N(N+1)(N+2)}\Bigl(S_{1}^2+S_{2}+2S_{-2}\Bigr)
              +\frac{8(5N^4+20N^3-N^2-14N+20)S_{1}}{9N(N+1)^2(N+2)^2}
 \N\\ && \hspace{-10mm}
              -\frac{2P_8}{9(N-1)N^3(N+1)^3(N+2)^3}
           \Biggr)
  +2C_F^2T_F\Biggl(
               \frac{8(N^2+N+2)}{N(N+1)(N+2)}
                \Bigl(3S_{2}S_{1}-S_{1}^3+4S_{-2}S_{1}+2S_{3}-4S_{-2,1}\Bigr)
 \N\\ && \hspace{-10mm}
              -\frac{16(N^2+N+2)S_{-2}}{N^2(N+1)^2(N+2)}
              -\frac{6(N^2+N+2)(3N^2+3N+2)S_{2}}{N^2(N+1)^2(N+2)}
              +\frac{2(3N^4+14N^3+43N^2+48N+20)S_{1}^2}{N^2(N+1)^2(N+2)}
 \N\\ && \hspace{-10mm}
              -\frac{4P_9S_{1}}{N^3(N+1)^3(N+2)}
              +\frac{P_{10}}{2N^4(N+1)^4(N+2)}
              +\frac{16(N^2+N+2)S_{-3}}{N(N+1)(N+2)}
            \Biggr)
  +4C_FT_F^2\Biggl(
 \N\\ && \hspace{-10mm}
               \frac{4(N^2+N+2)}{3N(N+1)(N+2)}\Bigl(S_{2}-3S_{1}^2\Bigr)
              +\frac{8(5N^3+14N^2+37N+18)S_{1}}{9N^2(N+1)(N+2)}
              -\frac{P_{11}}{9(N-1)N^4(N+1)^4(N+2)^3}
            \Biggr)
 \N\\ && \hspace{-10mm}
  +2C_FC_AT_F\Biggl(
                \frac{4(N^2+N+2)}{N(N+1)(N+2)}
                 \Bigl(4S_{1}^3-2S_{3}+4S_{-2,1}-2S_{-3}-3S_{-2}\Bigr)
 \N\\ && \hspace{-10mm}
               +\frac{4P_{12}S_{1}^2}{3(N-1)N^2(N+1)^2(N+2)^2}
               -\frac{4P_{13}S_{1}}{9(N-1)N^3(N+1)^3(N+2)^3}
 \N\\ && \hspace{-10mm}
               +\frac{P_{14}}{18(N-1)N^3(N+1)^3(N+2)^3}
               +\frac{4(N^2+N+2)(N^4+2N^3+8N^2+7N+18)S_{2}}{3(N-1)N^2(N+1)^2(N+2)^2}
            \Biggr)~.\label{AQg32}
 \end{eqnarray}
 }
%
%
{\small
 \begin{eqnarray}
  a_{Qg}^{(3),1}&=&
    \frac{1}{2}\hat{\gamma}_{qg}^{(2)}(n_f)
   -\frac{n_f}{2}\hat{\tilde{\gamma}}_{qg}^{(2)}(n_f)
   +4T_F^2n_f\Biggl(
       C_F\Biggl(
             \frac{4(N^2+N+2)}{9N(N+1)(N+2)}
              \Bigl(4S_{3}-S_{1}^3-3S_{2}S_{1}\Bigr)
            +\frac{4(3N+2)S_{1}^2}{3N^2(N+2)}
 \N\\ && \hspace{-10mm}
            +\frac{4(N^4-N^3-20N^2-10N-4)S_{1}}{3N^2(N+1)^2(N+2)}
            +\frac{2P_{15}}{3(N-1)N^5(N+1)^5(N+2)^4}
            +\frac{4P_{16}S_{2}}{3(N-1)N^3(N+1)^3(N+2)^2}
          \Biggr)
 \N\\ && \hspace{-10mm}
      +C_A\Biggl(
             \frac{4(N^2+N+2)}{9N(N+1)(N+2)}
              \Bigl( S_{1}^3+9S_{2}S_{1}+6S_{-3}+12S_{-2}S_{1}+8S_{3}
                    -12S_{-2,1}\Bigr)
            -\frac{4P_{17}S_{1}}{3N(N+1)^3(N+2)^3}
 \N\\ && \hspace{-10mm}
            -\frac{4(N^3+8N^2+11N+2)S_{1}^2}{3N(N+1)^2(N+2)^2}
            +\frac{4P_{18}}{3(N-1)N^4(N+1)^4(N+2)^4}
            -\frac{4P_{19}S_{2}}{3(N-1)N^2(N+1)^2(N+2)^2}
 \N\\ && \hspace{-10mm}
            +\frac{16(N^2-N-4)S_{-2}}{3(N+1)^2(N+2)^2}
          \Biggr)
           \Biggr)
   +2C_A^2T_F\Biggl(
                \frac{8(N^2+N+2)}{3N(N+1)(N+2)}
                  \Bigl( 12S_{-2,1}S_{1}-S_{1}^4-9S_{2}S_{1}^2-8S_{3}S_{1}
                        -6S_{-3}S_{1}
 \N\\ && \hspace{-10mm}
                        -12S_{-2}S_{1}^2\Bigr)
               -\frac{2P_{20}S_{1}^3}{9(N-1)N^2(N+1)^2(N+2)^2}
               +\frac{2P_{21}S_{1}^2}{3(N-1)N^2(N+1)^3(N+2)^3}
 \N\\ && \hspace{-10mm}
               -\frac{2P_{22}S_{1}}{3(N-1)N^4(N+1)^4(N+2)^4}
               -\frac{2P_{23}S_{2}S_{1}}{(N-1)N^2(N+1)^2(N+2)^2}
               -\frac{2P_{24}}{3(N-1)^2N^5(N+1)^5(N+2)^5}
 \N\\ && \hspace{-10mm}
               +\frac{4(N^2+N+2)(11N^4+22N^3-35N^2-46N-24)}{9(N-1)N^2(N+1)^2(N+2)^2}\Bigl(6S_{-2,1}-4S_{3}-3S_{-3}\Bigr)
 \N\\ && \hspace{-10mm}
               -\frac{8(N^2-N-4)(11N^4+22N^3-35N^2-46N-24)S_{-2}}{3(N-1)N(N+1)^3(N+2)^3}
               -\frac{8P_{25}S_{-2}S_{1}}{3(N-1)N^2(N+1)(N+2)^2}
 \N\\ && \hspace{-10mm}
               +\frac{2(11N^4+22N^3-35N^2-46N-24)P_{19}S_{2}}{3(N-1)^2N^3(N+1)^3(N+2)^3}
             \Biggr)
   +4C_AT_F^2\Biggl(
                \frac{8(N^2+N+2)}{9N(N+1)(N+2)}
                  \Bigl( S_{1}^3+8S_{3}-12S_{-2,1}
 \N\\ && \hspace{-10mm}
                        +9S_{2}S_{1}+12S_{-2}S_{1}+6S_{-3}\Bigr)
               -\frac{8(N^3+8N^2+11N+2)S_{1}^2}{3N(N+1)^2(N+2)^2}
               -\frac{8P_{19}S_{2}}{3(N-1)N^2(N+1)^2(N+2)^2}
 \N\\ && \hspace{-10mm}
               +\frac{2P_{26}}{27(N-1)N^4(N+1)^4(N+2)^4}
               +\frac{32(N^2-N-4)S_{-2}}{3(N+1)^2(N+2)^2}
               +\frac{4P_{27}S_{1}}{27N(N+1)^3(N+2)^3}
             \Biggr)
   +2C_F^2T_F\Biggl(
 \N\\ && \hspace{-10mm}
                \frac{8(N^2+N+2)}{3N(N+1)(N+2)}
                 \Bigl(4S_{3}S_{1}-S_{1}^4-3S_{2}S_{1}^2\Bigr)
               +\frac{2(3N^4+42N^3+107N^2+92N+28)S_{1}^3}{3N^2(N+1)^2(N+2)}
 \N\\ && \hspace{-10mm}
               +\frac{2P_{28}S_{1}^2}{N^3(N+1)^2(N+2)}
               +\frac{2P_{29}S_{1}}{N^4(N+1)^4(N+2)}
               +\frac{2(7N^4+74N^3+79N^2-12N-4)S_{2}S_{1}}{N^2(N+1)^2(N+2)}
 \N\\ && \hspace{-10mm}
               -\frac{8(N^2+N+2)(3N^2+3N+2)S_{3}}{3N^2(N+1)^2(N+2)}
               -\frac{2(3N^2+3N+2)(N^4+17N^3+17N^2-5N-2)S_{2}}{N^3(N+1)^3(N+2)}
 \N\\ && \hspace{-10mm}
               -\frac{P_{30}}{N^5(N+1)^5(N+2)}
            \Biggr)
   +4C_FT_F^2\Biggl(
                \frac{8(N^2+N+2)}{9N(N+1)(N+2)}
                 \Bigl(4S_{3}-S_{1}^3-3S_{2}S_{1}\Bigr)
 \N\\ && \hspace{-10mm}
               +\frac{P_{31}}{3(N-1)N^5(N+1)^5(N+2)^2}
               +\frac{8(N^4-N^3-20N^2-10N-4)S_{1}}{3N^2(N+1)^2(N+2)}
               +\frac{8(N^4+17N^3+17N^2-5N-2)S_{2}}{3N^2(N+1)^2(N+2)}
\N\\ && \hspace{-10mm}
                +\frac{8(3N+2)S_{1}^2}{3N^2(N+2)}
             \Biggl)
   +2C_FC_AT_F\Biggl(
                 \frac{8(N^2+N+2)}{3N(N+1)(N+2)}
                  \Bigl( 2S_{1}^4+12S_{2}S_{1}^2+4S_{3}S_{1}-12S_{-2,1}S_{1}
                        +6S_{-3}S_{1}
\N\\ && \hspace{-10mm}
                        +12S_{-2}S_{1}^2\Bigr)
                +\frac{4P_{32}S_{1}^3}{9(N-1)N^2(N+1)^2(N+2)^2}
                -\frac{4P_{33}S_{1}^2}{3(N-1)N^3(N+1)^3(N+2)^3}
\N\\ && \hspace{-10mm}
                -\frac{4P_{34}S_{1}}{3(N-1)N^4(N+1)^4(N+2)^4}
                -\frac{4P_{35}S_{2}S_{1}}{3(N-1)N^2(N+1)^2(N+2)^2}
                -\frac{P_{36}}{3(N-1)N^4(N+1)^5(N+2)^4}
\N\\ && \hspace{-10mm}
                -\frac{8P_{37}S_{-2}S_{1}}{N^2(N+1)^2(N+2)^2}
                +\frac{4(N^2+N+2)(3N^2+3N+2)}{N^2(N+1)^2(N+2)}
                  \Bigl(2S_{-2,1}-S_{-3}\Bigr)
\N\\ && \hspace{-10mm}
                -\frac{8(N^2-N-4)(3N^2+3N+2)S_{-2}}{N(N+1)^3(N+2)^2}
                -\frac{2P_{38}S_{2}}{3(N-1)N^3(N+1)^3(N+2)^2}
\N\\ && \hspace{-10mm}
                -\frac{8(N^2+N+2)(29N^4+58N^3-41N^2-70N-48)S_{3}}{9(N-1)N^2(N+1)^2(N+2)^2}
            \Biggr)~.\label{AQg31}
 \end{eqnarray}
 }

\normalsize
\noindent
Here, $S_{\vec{a}} \equiv S_{\vec{a}}(N)$, $P_k$ denote some polynomials in $N$, 
cf.~\cite{BBK1}, and $\gamma_{ij}^{(2)}$ are the 3-loop anomalous dimensions. 
The expansion coefficients given above depend on harmonic sums up weight {\sf w = 3}. 
Numerical studies show, that within the kinematic region of HERA the constant
terms to (\ref{eq:log}) are as important as the logarithmic contributions. 
Further details will be given in \cite{BBK1}.

\end{document}